\documentclass[conference]{IEEEtran}

\usepackage{graphicx} \usepackage{epstopdf}  \usepackage{amssymb} \usepackage{amsmath} \usepackage{amsthm} \usepackage{algorithm} \usepackage{algorithmic} \usepackage{subfigure} \usepackage{url}

\newtheorem{definition}{Definition} \newtheorem{theorem}{Theorem}  \newtheorem{corollary}{Corollary}

\graphicspath{{./}}

\begin{document}

\title{Randomization Improving Online Time-Sensitive Revenue Maximization for Green Data Centers}

\author{\IEEEauthorblockN{Huangxin Wang}
\IEEEauthorblockA{
George Mason University\\
Fairfax, VA 22030\\
hwang14@gmu.edu}
\and
\IEEEauthorblockN{Jean X. Zhang}
\IEEEauthorblockA{
Virginia Commonwealth University\\
Richmond, VA 23284\\
jxzhang@vcu.edu}
\and
\IEEEauthorblockN{Bo Yang}
\IEEEauthorblockA{
Jilin University\\
Changchun, China 130012\\
ybo@jlu.edu.cn}
\and
\IEEEauthorblockN{Fei Li}
\IEEEauthorblockA{
George Mason University\\
Fairfax, VA 22030\\
fli4@gmu.edu}
}
\maketitle


\begin{abstract}
Green data centers have become more and more popular recently due to their sustainability. The resource management module within a green data center, which is in charge of dispatching jobs and scheduling energy, becomes especially critical as it directly affects a center's profit and sustainability. The thrust of managing a green data center's machine and energy resources lies at the uncertainty of incoming job requests and future showing-up green energy supplies. Thus, the decision of scheduling resources has to be made in an online manner. Some heuristic deterministic online algorithms have been proposed in recent literature. In this paper, we consider online algorithms for green data centers and introduce a randomized solution with the objective of maximizing net profit. Competitive analysis is employed to measure online algorithms' theoretical performance. Our algorithm is theoretical-sound and it outperforms the previously known deterministic algorithms in many settings using real traces. To complement our study, optimal offline algorithms are also designed.
\end{abstract}


\section{Introduction}

In this paper, we study the problem of scheduling jobs and energy in data centers. A \emph{data center} is a computing facility used to house computing systems and their associated components such as communication and storage subsystems. Usually, a data center stores data and provides computing functionalities to its customers. Through charging fees for data access and server services, a data center gains revenue~\cite{amazonprice}. At the same time, to maintain its running structure, a data center has to pay \emph{operational costs} including hardware costs (for example, those of upgrading computing and storage devices and air conditioning facilities), electrical bills for power supply, network connection costs, and personnel costs. To maximize a data center's net profit, we expect to increase the revenue gathered and simultaneously decrease the operational costs paid.

Unfortunately, the ever increasing power costs and energy consumption in data centers have brought many serious economic and environmental problems to our society and evoked significant attention recently. As reported, the estimates of annual power costs for U.S. data centers in 2010 reached as high as $3.3$ billion dollars~\cite{EnergyCost}. As a concrete example, in a modern high-scale data center with 45,000 to 50,000 servers, more than 70\% of its operational cost (around half a billion dollars per year)~\cite{usage} goes to maintaining the servers and providing power supply. The energy spending in data centers in 2014 is \$143 billion and is at a growth rate of nearly 7\%~\cite{inforweek}. Targeting on both economic and environmental factors, academic researchers and industrial policy makers have put a lot of effort in investigating engineering solutions to make data centers work better without sacrificing service qualities and environment sustainability.

A growing trend of reducing energy costs as well as protecting our clean environments is to fuel a data center using renewable energy from wind and solar power. We term this type of energy as ``\emph{green energy}'' as it comes from renewable and non-polluting sources. The amount and availability of green energy are usually intermittent and cannot be fully predicted in the long term. Another type of energy, called ``\emph{brown energy}'', comes from the available electrical grid in which the power is produced by carbon-intensive means. Brown energy's sources are much more stable and predictable. A data center with both green and brown energy supplies is called a \emph{green data center}. A natural goal of managing energy resources is to reduce the usage of brown energy if possible, while to maintain the levels of service quality to jobs.

In this paper, we design job and energy scheduling algorithms for green data centers. The ultimate goal is to optimize green and brown energy usage without sacrificing service qualities. Our research is built upon the work done by Goiri~\emph{et. al}~\cite{GoiriL11}, by Keskinocak~\emph{et. al}~\cite{KeskinocakRT01}, and by Bansal~\emph{et. al}~\cite{BansalCP11}. Within this framework, job requests arrive at a data center over time. An algorithm is to determine \emph{whether} (job admission), \emph{when} (job processing-window) and \emph{where} (job-machine matching) to schedule a job, as well as which type of energy to use in a time slot.  Note that different ways of assigning jobs to machines and different time slots, and feeding machines using different types of energy may result in different revenue and operational cost. Define \emph{net profit} as the difference between revenue and operational cost. We address the following question: \emph{How do we dispatch jobs and schedule green/brown energy to maximize net profit?} Recall that the information on later released jobs and future accessible green energy is in general unknown at the moment when the current scheduling decision is made, and thus, what we study in this paper can be regarded as an online version of a machine scheduling problem.

To evaluate an online algorithm's performance, we start from two perspectives. In theory, we use \emph{competitive ratio}~\cite{BorodinE98} to measure an online algorithm's worst-case performance against a clairvoyant adversary. Competitive analysis has been used widely to analyze online algorithms in computer science and operations research~\cite{BorodinE98}. In practice, we conduct simulations using both real traces and simulated data. The crux of our algorithmic idea in this paper is to introduce `internal randomness' in scheduling energy and jobs. As what we will see in the remaining parts of this paper, `randomness' helps both theoretically and empirically, particularly in adversarial settings.


\subsection{Problem formulation}

A data center is regarded as a \emph{resource provider} which provides a set of sharable machines for its clients. The clients, regarded as \emph{resource consumers}, have their jobs processed and in turn, pay for the service they get. The data center's \emph{revenue management} module has the objective of maximizing its \emph{net profit}, defined as the difference between the revenue collected from the clients and the operational costs charged to maintain the computing and networking system. Here the operational costs do not include those for upgrading systems, paying personnel, or training operators. We model a data center's revenue management as a decision-making problem of scheduling jobs and energy. The components of a computing system within a data center is pictured in Figure~\ref{fig:system} and we will introduce each of them in details as shown below.

\begin{figure}[h!]
\center
\includegraphics[width=.3\textwidth]{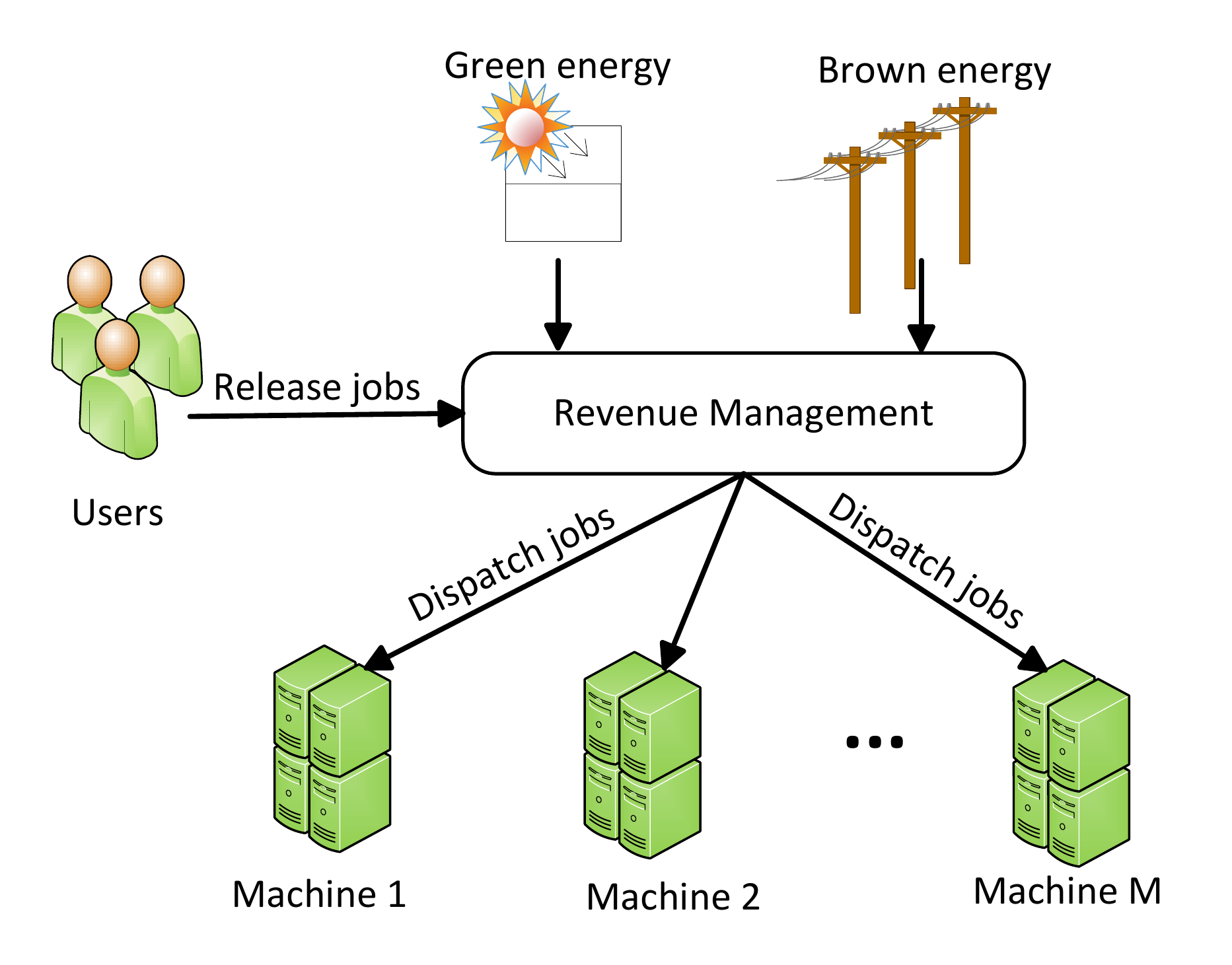}
\caption{Components of a solar-powered green data center}
\label{fig:system}
\end{figure}

We model a data center handling large batch-jobs, as what Facebook processes~\cite{SWIM, AdnanSGK12}. This model is the same as the one proposed in~\cite{PinheiroBCH01}. Some statistical data about machine settings~\cite{MohanS12} are shown in Table~\ref{tb:power_consumption} and Table~\ref{tb:setup-time}. Data centers usually have job processing time in the order of tens of minutes~\cite{KavulyaTGN10}. In the two real workloads traces in~\cite{Grid5k, Intrepid}, jobs have average processing time of $0.86$ and $1.44$ hours respectively, with medium being $12.01$ and $1.002$ hours respectively.

\begin{table}[h!]
\centering
\caption{Power consumption of physical server state}
\begin{tabular}{|l|l|}
\hline
state & power consumption (in Watts)\\ \hline \hline
BUSY & 240\\ \hline
IDLE & 150\\ \hline
SLEEP & 10\\ \hline
OFF & 0\\ \hline
\end{tabular}
\label{tb:power_consumption}
\end{table}

\begin{table}[ht!]
\centering
\caption{Time cost of physical server state transition}
\begin{tabular}{|l|l|}
\hline
from state & to IDLE state (in seconds)\\ \hline
SLEEP (Hibernate/Suspend) & 25\\ \hline
OFF & 48\\ \hline
IDLE & 0\\ \hline
\end{tabular}
\label{tb:setup-time}
\end{table}

\paragraph*{Machine resources}

Time is discrete. A data center hosts $M \in \mathbb{Z^+}$ \emph{machines} (also called \emph{nodes}) to schedule jobs. At any time, a machine can process at most one job. To make these machines function, electrical power resource is consumed at the time when jobs are being executed. We normalize the energy costs such that without loss of generality, we assume that a machine consumes $1$ unit of energy per time slot when it is processing a job and $0$ unit otherwise. This simplification is supported by the negligible machine transition time cost, compared with batch-job sizes.

\paragraph*{Job requests}

Clients release jobs to a data center to be processed. Jobs arrive over time. At a time, some (may be $0$) jobs arrive. Each job $j$ has an integer \emph{arriving time} (also called \emph{release time}) $r_j \in \mathbb{Z}^+$, an integer \emph{processing time} $p_j \in \mathbb{Z}^+$, an integer \emph{deadline} $d_j \in \mathbb{Z}^+$, and an integer \emph{machine requirement} $q_j \in [1, M]$. Running one job $j$ may require more than one machines to be simultaneously running at a time. The total \emph{machine resource requirement} for a job $j$ is defined as $q_j \times p_j$. The resource management module specifies whether and where to schedule a job upon its arrival. A successfully completed job $j$ needs to be executed in a consecutive time period without being interrupted, preempted, or migrated~\cite{GoiriL11}, starting at a time in-between $r_j$ and $d_j - p_j$. We can use a triple $(r_j, p_j, q_j)$ to denote a job $j$.

\paragraph*{Time-sensitive revenue}

A client pays to the data center for the service he receives. The payoff depends on the job's machine resource requirement as well as the service quality. Consider a job $j = (r_j, p_j, q_j)$. Let $s_j$ denote the \emph{starting time} to execute $j$ and $c_j$ ($c_j := s_j + p_j$) denote $j$'s \emph{completion time}~\cite{KargerWS10}. For continuous job streams, we revise the term \emph{stretch}~\cite{BenderCM98, BenderMR02} to characterize the \emph{service quality} $l_j$ that a job $j$ receives. (Recall that in cloud computing and data center services, a client assumes that he gets to be served immediately upon delivering his job request.) Define $l_j := \frac{p_j}{c_j - r_j}$, where $c_j \ge r_j + p_j$. Each client pays to the data center money (revenue) proportional to its machine resource consumed: $v_j :=
\begin{cases}
\$\beta \times p_j \times q_j, & \mbox{if } l_j \ge L_j\\
\$0, & \mbox{otherwise}
\end{cases}
$. The parameter $\beta$ is called \emph{service charging rate}~\cite{amazonprice} and $L_j$ ($L_j \in (0, 1]$) is the \emph{least service quality} that a client $j$ can receive.) As what is specified by Amazon EC2 data center service~\cite{amazonprice}, different clients pay various rates of service fee for per unit of different types of jobs to be processed. To ensure that we receive money $v_j$ from a job $j$, we need to guarantee $\frac{p_j}{c_j - r_j} \ge L_j$. Thus, we define $d_j$ as the \emph{deadline} of completing a job $j$, where $d_j := r_j + \frac{p_j}{L_j}$, to indicate the time by which the job $j$ should be completed to satisfy its quality requirement.

\paragraph*{Time-sensitive energy costs}

Energy is consumed along the course of machines executing jobs. Usually, a data center is able to predict green energy quantity only within a 48-hour \emph{scheduling window} (see~\cite{GoiriL11} and the references therein). Different types of energy costs vary over time. Spending green energy costs us nothing. Unfortunately, no batteries are used to store any surplus green energy~\cite{Bianchini12}, due to economic concerns and technical difficulties. The brown energy's unit-cost is \emph{time-sensitive} and thus it is a variable related to on-peak/off-peak time periods. A unit of brown energy has price $\$B^d$ when at on-peak (usually at daytime) and price $\$B^n$ when at off-peak (usually at nighttime). This assumption is the most common one used in modeling brown electricity pricing~\cite{GoiriL11}. For instance, the prices charged by an integrated generation and energy service company in New Jersey~\cite{GoiriL11} have $B^d = \$0.13/kWh$ and $B^n = \$0.08/kWh$.

\paragraph*{Objective}

Scheduling jobs successfully can earn a data center some \emph{revenue} and paying for any brown energy used (to power a data center, along with the limited green energy) incurs \emph{operational costs}. We define
\begin{displaymath}
\mbox{net profit = revenue - operational cost},
\end{displaymath}
where \emph{revenue} is the total money gained through finishing jobs and operational cost is the total brown energy cost that the service provider consumes to run the jobs. The objective of revenue management module of green data centers is to design a scheduler to complete all or a subset of the released jobs in order to maximize net profit. We call this problem GDC-RM, standing for `Green Data Center's Revenue Management'. In the remaining parts of this paper, we present combinatorial optimization algorithms for GDC-RM. Recall that job requests and energy arriving information are unknown beforehand, GDC-RM is essentially an online decision-making problem.


\subsection{Related work}

How to schedule green energy in an efficient and effective manner has been investigated extensively. Although green energy has the advantages of being cost-effective and environmental-friendly, there is a challenge in using it due to its daily seasonal variability. Another challenge comes from customers' workload fluctuations~\cite{HeY10}. There could lead to a temporal mismatch between the green energy supply and the workload's energy demand in the time axis --- a heavy workload arrives when the green energy supply is low. One solution is to ``bank'' green energy in batteries for later possible use. However, this approach incurs huge energy lost and high additional maintenance cost~\cite{Bianchini12}. Thus, a run-time online algorithm for a matching of workload and energy is highly demanded for green data centers.

Two green data center settings have been considered: (1) centralized data centers (such as in~\cite{GoiriL11, Krioukov11, LiQ11}) and (2) geographically distributed data centers (such as in~\cite{LiuL11, LinL12}). The objectives to optimize are usually classified as (a) to maximize green energy consumption, (b) to minimize brown energy cost, and (c) to maximize profits. In addition, some researchers incorporated dynamic pricing of brown energy in their models~\cite{GoiriL11, RaoL10, LiuLLW14}. Unlike the model studied in this paper, research on geographical data centers focuses on distributing workloads among distributed data centers in order to consume the available free green energy or relative cheaper brown energy at other data centers. Although geographical data centers have become popular nowadays for big companies such as Google and Amazon, small-scale centralized data centers are still important since as reported, numerous small and medium-sized companies are the main contributors to the energy consumed by data centers~\cite{epa}. There exists a huge impact in studying the problem of revenue management for centralized data centers.

Among the work on centralized data centers, ~\cite{GoiriL11, Krioukov11} studied a model which is the same as ours presented in this paper. \cite{AksanliV11} aimed to improve green energy usage and~\cite{LiuC12} had the goal of reducing brown energy costs. The algorithmic idea underlying the above-mentioned solutions is greedy and they employed algorithms known as \emph{First-Fit} and \emph{Best-Fit}. All prior work focuses on either maximizing green energy consumption or minimizing brown energy consumption/cost except~\cite{GhamkhariR13} which studied the net profit maximization problem for centralized data center service providers. ~\cite{GhamkhariR13} proposed a systematic approach to maximize green data center's profit with a stochastic assumption over the workload --- the workload that they studied is restricted to online service requests with variable arrival rates. In this paper, we study the profit maximization problem in a more general setting. In particular, we do not make \emph{any} particular assumptions over the workload's stochastic property. In addition, we incorporate dynamic brown energy price in our model which is a widely used energy charging scheme in data centers.


\section{Online Algorithms}

The offline version of GDC-RM is NP-hard which can be proved via a reduction to the well-known NP-hard Knapsack problem~\cite{GareyJ79} as shown in Appendix~\ref{appendix_GDC-RM_hardness}. In reality, job scheduling in data centers is essentially an online problem. For the online version of the problem GDC-RM, we first discuss two widely-used heuristic online algorithms First-Fit and Best-Fit and analyze their limitations. Then we propose a randomized algorithm Random-Fit.  We conduct \emph{competitive analysis} when we evaluate an online algorithm's theoretical performance. Competitive analysis is used to compare the output of an online algorithm with that of an optimal offline clairvoyant algorithm. This unrealistic offline algorithm is assumed to know all the input information (including the green energy arrivals and units, brown energy prices, and job arriving sequences) beforehand.

\begin{definition}[Competitive ratio~\cite{BorodinE98}]
A deterministic (respectively, randomized) online algorithm ON is called $k$-competitive if its (respectively, expected) performance of any instance is at least $1 / k$ times of that of an optimal offline algorithm. The optimal offline algorithm is also called (respectively, oblivious) adversary. Let OPT denote the optimal offline solution of an input. Competitive ratio $k$ is defined as $k := \max \limits_{I} \frac{OPT - \delta}{E[ON]}$, where $\delta$ is a constant and $E[ON]$ is ON's (expected) output of an input.
\end{definition}

Note that unlike stochastic algorithms which heavily rely on the statistical assumptions on the input sequence, competitive online algorithms guarantee the worst-case performance in any given finite time frame against its adversary. The workload (input) does not need to satisfy any stochastic assumptions. Competitive analysis is used when rigorous analysis of online algorithms is needed and when the input's stochastic properties are hard to get. For the problem GDC-RM, a green data center's workloads are difficult to model~\cite{MeisnerW10} and thus competitive analysis acts as a suitable metric.


\subsection{Competitive analysis of First-Fit, Best-Fit, and GreenSlot}

First-Fit is a conventional deterministic online scheduler which schedules, if possible, a job to the earliest available time slots regardless of its energy cost. Although this approach can cause minimum delay of a job and maximum throughput, it might not achieve a good overall profit due to high brown energy cost needed to finish a job in earlier time slots (instead of using green energy or less-expensive night-time brown energy in later time slots).

Best-Fit is also a widely-used heuristic and deterministic online algorithm. Actually, First-Fit and Best-Fit have been used extensively in online 1D bin-packing~\cite{CoffmanCGMV13} and 2D bin-packing problems~\cite{LodiMM02}. The Best-Fit algorithm locates the most `\emph{cost-efficient}' time slots to schedule a job. It picks up the \emph{best} time interval to schedule a job in a myopic way and it does not take later job arrivals or energy supplies into account. As pointed out in~\cite{GoiriL11}, Best-Fit may reject more jobs or miss more deadlines than First-Fit does. The reason lies at the observation that Best-Fit always delays jobs to the best cost-efficient time slots regardless of future workload for those time slots. As a result, some jobs may fail to be scheduled due to deadline constraints and thus the profit is harmed. GreenSlot~\cite{GoiriL11} is a variant of Best-Fit; a heuristic modification is made to avoid rejecting future-arriving jobs due to delaying scheduling current jobs. GreenSlot adds a penalty in postponing scheduling jobs at time slots that are likely to cause a job to miss its deadline. This penalty is a manly-tuned-up parameter to fit various job sets and thus it is workload dependent and cannot guarantee to improve the worst case profits nor to be used as a universal algorithm handling various job requests.

We analyze the competitive ratio of First-Fit and Best-Fit in the following. First we introduce some notations appearing in Theorem~\ref{comRatio_FF}, ~\ref{comRatio_BF} and ~\ref{comRatio_RF}. In specific, we normalize the costs of green energy and brown energy. According to the definition of profit, a job $j$ with $p_j$ processing time and $q_j$ machine requirement has profit $c \cdot p_j \cdot q_j - \int_t P(t)$, where $P(t)$ has the value $0$ (for green energy), $B^d$ (for on-peak brown energy), or $B^n$ (for off-peak brown energy) respectively when the job is processed using various types of energy. $P(t)$ is in integral along the time when the machines process $j$. If all jobs are with the same processing time and machine requirements, then we normalize the profit as $\frac{c \cdot p_j \cdot q_j - \int_t P(t)}{c \cdot p_j \cdot q_j} = 1 - \int_t \frac{P(t)}{c \cdot p_j \cdot q_j}$. In our proofs below, we generate instances such that for each job, it is processed by \emph{only} one type of energy using the particular algorithm. Thus, for ease to present the competitive ratio, we define $1 - \frac{P(t)}{c \cdot p_j \cdot q_j}$ as $v_{on}$, $v_{off}$, $v_g$ as below.

\begin{align*}
& 1 - \frac{P(t)}{c \cdot p_j \cdot q_j} := \\
& \begin{cases}
v_{on}, & \text{if only use on-peak brown energy to schedule $j$}\\
v_{off}, & \text{if only use off-peak brown energy to schedule $j$}\\
v_g, & \text{if only use green energy to schedule $j$}
\end{cases}
\end{align*}

Note that the normalized profit $1 - \frac{P(t)}{c \cdot p_j \cdot q_j}$ has a value among $(0, 1]$. According to the fact that on-peak brown energy is expensive than off-peak brown energy. Also, green energy has cost $0$. We have $0 < v_{on} < v_{off} < v_g = 1$. Also, for jobs with the same processing times and same machine requirements, they have the same value for $v_{on}$, $v_{off}$, and $v_g$.

\begin{theorem}
The lower bound of competitive ratio for First-Fit is $\max \left\lbrace\frac{v_{off}}{v_{on}}, \frac{v_g}{v_{on}}\right\rbrace$.
\label{comRatio_FF}
\end{theorem}

\begin{IEEEproof}
To prove the lower bound, we create an input instance. Let OPT denote an optimal offline algorithm. We assume each job has processing time requirement $p_j = 1$ and machine requirement $q_j = M$. We use $(r, d)$ to denote a job with release time $r$ and deadline $d$. We have $M$ machines.

Assume there are two daytime time slots $t_1$ and $t_2$, with $0$ and $M$ green energy units arriving at them respectively. Assume there is only one job $j = (t_1, t_2)$ arriving. First-Fit schedules $j$ at time $t_1$, earning a revenue $v_{on}$. OPT schedules $j$ at time $t_2$, achieving a profit $v_g$. The competitive ratio is $\frac{OPT}{FF} = \frac{v_g}{v_{on}}$.

If $t_1$ is at on-peak and $t_2$ is at off-peak, then we assume that no green energy arrives at both time slots. Using the same analysis approach, we get the competitive ratio $\frac{OPT}{FF} = \frac{v_{off}}{v_{on}}$. Therefore, we conclude that First-Fit has a competitive ratio at least $\max\left\lbrace \frac{v_{off}}{v_{on}}, \frac{v_g}{v_{on}}\right\rbrace$.
\end{IEEEproof}

\begin{theorem}
The lower bound of competitive ratio for Best-Fit is $\max\left\lbrace 1+\frac{v_{on}}{v_{off}}, 1 + \frac{v_{off}}{v_g}\right\rbrace$.
\label{comRatio_BF}
\end{theorem}

\begin{IEEEproof}
We prove via constructing an input instance as a lower bound example. We assume all the arriving jobs have processing time $p_j = 1$ and machine requirement $M$ (no two jobs can be executed simultaneously at the same time slot).

Assume there are two time slots $t_1$ and $t_2$ --- $t_1$ is at on-peak while $t_2$ is at off-peak. There are no green energy arriving at both time slots. Assume there are two jobs released $j_1 = (t_1, t_2)$ and $j_2 = (t_2, t_2)$.

Best-Fit will delay job $j_1$ to be scheduled at time $t_2$, resulting in a deadline conflict between jobs $j_1$ and $j_2$, and thus only gain profit $v_{off}$. OPT will schedule $j_1$ and $j_2$ at time $t_1$ and $t_2$ respectively, gaining a profit $v_{on} + v_{off}$. Thus the competitive ratio is $\frac{OPT}{BF} = 1 + \frac{v_{on}}{v_{off}}$.

If $t_1$ is at off-peak and $t_2$ is at on-peak, then we assume there are $0$ and $M$ units of green energy arrive at time $t_1$ and $t_2$ respectively. Using the same analysis approach, we get the competitive ratio $\frac{OPT}{BF} = 1 + \frac{v_{off}}{v_g}$. we conclude that Best-Fit has a competitive ratio at least $\max\left\lbrace 1 + \frac{v_{on}}{v_{off}}, 1 + \frac{v_{off}}{v_g}\right\rbrace$.
\end{IEEEproof}

Based on the above analysis and recall $0 < v_{on} < v_{off} < v_g = 1$, we have the following result.

\begin{corollary}
Deterministic algorithms First-Fit and Best-Fit, with or without job preemption, have competitive ratios no strictly better than $2$, even for a restricted case in which all jobs are with the same length.
\end{corollary}

As shown above, First-Fit and Best-Fit have arbitrary worse competitive ratios. Even for the special case in which all jobs are with the same processing times and the same node requirements, First-Fit and Best-Fit have competitive ratios no better than $2$. The crux of competitive analysis lies as below: On one hand, if we schedule a job regardless of its alone energy cost, then we favor the algorithm First-Fit as it leaves room for later arriving jobs to be scheduled. On the other hand, if we schedule a job considering its energy cost, then this job may be scheduled at a later time when energy is cheaper (e.g., the green energy runs out for now and there exists predicted green energy in the future). The potential risk is that Best-Fit may prevent admitting and completing a later released job. Such a deterministic online algorithm is then pessimistic in terms of competitive ratio.


\subsection{Randomized algorithm Random-Fit}
\label{alg:rf}

In order to solve this dilemma introduced by First-Fit and Best-Fit in maximizing net profit, we introduce an algorithm with internal randomness to twist the high brown energy cost that we pay right now and the high cost of losing potential future jobs. We remark here that the randomness (i.e., the probability $p$ in the following Algorithm~\ref{alg_Random-Fit}) does not depend on the job workload at all and thus, we do not have to derive $p$ from any stochastic features assumed from the input. We develop an algorithm called Random-Fit (as in Algorithm~\ref{alg_Random-Fit}) with parameter $p$ and we will show how to set the value of $p$ to lead to the optimal result.

\begin{algorithm}
\caption{Random-Fit (RF)}
\begin{algorithmic}[1]
\STATE Let $j$ denote an arriving job.

\IF{there is sufficient free green energy to schedule $j$}

\STATE schedule $j$ at its earliest time interval;

\ELSE

\STATE use probability $p$ to schedule $j$ at its earliest time interval;

\STATE use probability $1 - p$ to schedule $j$ at its most economic time interval.

\ENDIF
\end{algorithmic}
\label{alg_Random-Fit}
\end{algorithm}

In the following, we consider a special case in which all jobs are with the same lengths and the same node requirements. We calculate the \emph{optimal} value for $p$ in the following analysis.

\begin{theorem}
In scheduling jobs with the same processing times and the same node requirements, algorithm Random-Fit has its competitive ratio $c = \max \left\lbrace 1 + \frac{v_{on}}{v_{off}} - \left(\frac{v_{on}}{v_{off}}\right)^2, 1 + \frac{v_{off}}{v_g} - \left(\frac{v_{off}}{v_g}\right)^2 \right\rbrace$, against an oblivious adversary. This competitive ratio $c$ is no more than $1.25$.
\label{comRatio_RF}
\end{theorem}

\begin{IEEEproof}
Let OPT denote an optimal offline algorithm (an oblivious adversary) as well as its net profit. Let RF denote the Random-Fit algorithm as well as its expected net profit. In order to prove this theorem, we will show that $\frac{OPT}{RF} \le 1.25$.

We employ a charging scheme to prove Theorem~\ref{comRatio_RF}. Initially, OPT and RF have the same energy resource and machine resource. We consider an arriving job at time $t_1$ with the inductive assumption that before time $t_1$, the ratio between OPT and RF is no more than $c$ (in Theorem~\ref{comRatio_BF}, $c = 1.25$). In the following, we show that after time $t_1$, the inductive assumption still holds.

Two facts are used in the proof: (1) Randomness only plays its role when no green energy is available (otherwise, no random decision is needed, see Algorithm~\ref{alg_Random-Fit}); and (2) If OPT schedules a job at time $t$, then OPT schedules the earliest-deadline job as all jobs are with the same processing times and node requirements. We will show that the following invariant holds: At any time, the net profit ratio between OPT and RF is no more than $c$; also, OPT has no more remaining green energy than RF does, if we charge appropriate revenue to OPT. This includes the scenario in which OPT schedules a job later with energy consumption while we charge the revenue and the energy cost for now for OPT. Once this invariant holds, Theorem~\ref{comRatio_RF} holds immediately. We consider the release jobs via case study and use $(r, d)$ to denote a job with release time $r$ and deadline $d$.

\paragraph{Consider the two neighboring time slots $t_1$ and $t_2$ which are at on-peak and at off-peak respectively}

\begin{enumerate}
\item OPT releases one job $j_1 = (t_1, t_2)$ and OPT schedules $j_1$ at time $t_2$, achieving a profit $v_{off}$. While RF will schedule job $j_1$ to time $t_1$ with probability $p$ and to time slot $t_2$ with probability $1 - p$, earning an expected profit $p \cdot v_{on} + (1 - p) \cdot v_{off}$. In this case, the competitive ratio is $\frac{OPT}{RF} = \frac{v_{off}}{p \cdot v_{on} + (1 - p) \cdot v_{off}}$.

\item OPT releases two jobs $j_1 = (t_1, t_2)$ and $j_2 = (t_2, t_2)$. OPT would schedule $j_1$ at time $t_1$ and schedule $j_2$ at time $t_2$, achieving a profit of $v_{on} + v_{off}$. While, the RF will schedule $j_1$ at time $t_1$ with probability $p$ and schedule either $j_1$ or $j_2$ at time $t_2$ (due to the job deadline constraints), earning a profit of $p \cdot v_{on} + v_{off}$. Therefore, the competitive ratio is $\frac{OPT}{RF} = \frac{v_{on} + v_{off}}{p \cdot v_{on} + v_{off}}$.
\end{enumerate}

In this scenario, the competitive ratio is: $\min_p \left\{\max\left\{\frac{v_{off}}{p \cdot v_{on} + (1 - p) \cdot v_{off}}, \frac{v_{on} + v_{off} }{p \cdot v_{on} + v_{off}}\right\}\right\}$. In solving above min-max problem, $p = \frac{x}{1 + x - x^2}$ where $x = \frac{v_{on}}{v_{off}}$ optimizes the competitive ratio $\frac{OPT}{RF} = 1 + x - x^2 = 1 + \frac{v_{on}}{v_{off}} - \left(\frac{v_{on}}{v_{off}}\right)^2 \le 1.25$.

\paragraph{Consider the two neighboring time slots $t_1$ and $t_2$ which are at off-peak and at on-peak respectively}

\begin{enumerate}
\item OPT releases one job $j_1 = (t_1, t_2)$. The worst-case is that OPT uses the on-peak day's free green energy to schedule this job $j_1$. Using the same analysis approach, we get a competitive ratio $\frac{OPT}{RF} = \frac{v_g}{p \cdot v_{off} + (1 - p) \cdot v_g}$.

\item OPT releases two jobs $j_1 = (t_1, t_2)$ and $j_2 = (t_2, t_2)$. Similarly, we get a competitive ratio $\frac{OPT}{RF} = \frac{v_{off} + v_g}{p \cdot v_{off} + v_g}$.
\end{enumerate}

In this scenario, the competitive ratio is $\min_p \left\lbrace\max\left\lbrace \frac{v_g}{p \cdot v_{off} + (1 - p) \cdot v_g} , \frac{v_g + v_{off} }{p \cdot v_{off} + v_g}\right\rbrace \right\rbrace$. Similarly, when $p = \frac{y}{1 + y - y^2}$ where $y = \frac{v_{off}}{v_g}$ ($0 < y < 1$), we get the optimal competitive ratio $\frac{OPT}{RF} = 1 + y - y^2 = 1 + \frac{v_{off}}{v_g} - \left(\frac{v_{off}}{v_g}\right)^2  \leq 1.25$.
\end{IEEEproof}

\begin{corollary}
Random-Fit has a better competitive ratio compared to First-Fit and Best-Fit.
\end{corollary}

\begin{IEEEproof}
In Theorem~\ref{comRatio_FF} and Theorem~\ref{comRatio_BF}, the lower bounds of competitive ratio for First-Fit and Best-Fit are proved to be $\max \left\lbrace\frac{v_{off}}{v_{on}}, \frac{v_g}{v_{on}}\right\rbrace$ and $\max\left\lbrace 1 + \frac{v_{on}}{v_{off}}, 1 + \frac{v_{off}}{v_g}\right\rbrace$ respectively.
We use FF and BF to stand for First-Fit and Best-Fit. As the expected competitive ratio of Random-Fit is no larger than $\max\left\{1 + \frac{v_{on}}{v_{off}} - \left(\frac{v_{on}}{v_{off}}\right)^2, 1 + \frac{v_{off}}{v_g} - \left(\frac{v_{off}}{v_g}\right)^2\right\}$, we have $\frac{OPT}{RF} < \frac{OPT}{BF}$. Since $1 + k - k^2 < 1 / k$ (where $0 < k <1$), we have $1 + \frac{v_{on}}{v_{off}} - \left(\frac{v_{on}}{v_{off}}\right)^{2} < \frac{v_{off}}{v_{on}}$, and $1 + \frac{v_{off}}{v_g} - \left(\frac{v_{off}}{v_g}\right)^2  < \frac{v_g}{v_{off}}$, then we have $\frac{OPT}{RF} < \frac{OPT}{FF}$.
\end{IEEEproof}

\begin{corollary}
The optimal randomness (probability) for Random-Fit is $\begin{cases}
p = \frac{x}{1 + x - x^2}, & x = \frac{v_{on}}{v_{off}}\\
p' = \frac{y}{1 + y - y^2}, & y = \frac{v_{off}}{v_g}
\end{cases}$ for scheduling jobs from on-peak time to off-peak time and from off-peak time to on-peak time respectively.
\end{corollary}

We remark here that using Yao's principle~\cite{Yao77, BorodinE98}, the lower bound of online randomized algorithms on the general case can be easily derived from our competitive analysis on deterministic algorithms.


\subsection{On offline algorithms and resource augmentation approach for analyzing online algorithms}

An optimal offline algorithm for the cases in which jobs have same processing times and node requirements can be formulated using a linear program.

Consider an online algorithm and its competitive analysis, as what we have done in Section~\ref{alg:rf}. Completive ratio is commonly used to measure an online algorithm's performance degradation when the future input information is totally unknown. A theoretical measure called \emph{resource augmentation} was introduced, especially for analyzing online algorithms with poor worst-case performance~\cite{KalyanasundaramP00, PruhsS10, PhillipsSTW02}. The underlying idea is to increase the scarce resource to compensate the loss due to limited knowledge about future.

An online algorithm ALG is called $\alpha$-resource $c$-competitive if the online algorithm is given $\alpha$ times of resources while the adversary has only one unit of resource, and ALG's performance is no worse than $1 / c$ times of what ADV achieves. Evidence has shown that if $\alpha$ is small while $c$ decreases significantly, then the algorithm ALG has much better practical performance than what its theoretical bounds show. In the problem GDC-RM, green energy is a kind of scarce resource. We have the following negative result on the resource augmentation approach.

\begin{theorem}
The lower bound of competitive ratio for $\alpha$-times green energy augmentation is no better than $\max\{ \frac{v_g}{v_{on}}, 1 + \frac{v_{on}}{v_{off}}, 1 + \frac{v_{off}}{v_g}\}$.
\label{thm:reau}
\end{theorem}

The formulation of the optimal offline algorithm is shown in Appendix~\ref{Appendix_offline}.


\section{Performance Evaluation}

In this section, we experimentally evaluate our designed randomized algorithm against GreenSlot~\cite{GoiriL11} scheduler as well as the two well-used deterministic online algorithms First-Fit and Best-Fit which have been adopted with possible tuning-ups in previous literature. An optimal offline algorithm is also developed, although its running time is tedious when the input size is large. The optimal offline algorithm is formulated as a binary integer program. We will first introduce the simulation setting, and then explain the simulation methodology, and finally report the simulation results and analysis.


\subsection{Simulation settings}

\paragraph*{Data center}

The simulated green data center is configured similar to the one in~\cite{GoiriL11} but with more machines (nodes). The data center is a cluster consisting of $100$ machines with each machine consumes $140$W when they are running jobs. The total energy consumption is the sum of the energy consumed by the machines when they are processing jobs over time.

\paragraph*{Green energy}

We use the solar energy trace from the Computer Science Weather Station at University of Massachusetts, Amherst~\cite{solarTrace}. The solar energy is fine grained such that it is collected every $5$ minutes. We scale down the solar energy trace and make it compatible with the simulated data center by making the peak solar power cover the maximum possible power consumption. We select an arbitrary $5$-day-period time to simulate the solar energy trace input.





\paragraph*{Brown energy price}

The brown energy price is varying at on-peak/off-peak periods. The electricity cost is less at off-peak and more at on-peak periods. We use the prices charged by PSEG in New Jersey at summer time~\cite{GoiriL11} as an example: on-peak price (from 9am to 11pm) $0.13/kWh$, off-peak price (from 11pm to 9 am) $0.08/kWh$.

\paragraph*{Service pricing}

The green data center service provider charges the clients for the computing resource consumed. We set the service price based on Amazon EC2' pricing~\cite{amazonprice}. The charging price is set as $\$0.022/h$ per machine.

\paragraph*{Workloads}

We use real workload traces \emph{Grid5k} as the workload input in our simulation. Grid5k~\cite{Grid5k} is a real workload trace which was collected from Grid'5000 system~\cite{Grid5000Platform}, a-$2218$ node experimental grid platform consisting of 9 sites geographically distributed in France, from May 2004 to November 2006.

We randomly select 2 five-day-period workloads, denoted as \emph{Grid5k-1} and \emph{Grid5k-2} as the workload input in the simulation. Note that in order to simulate various workload utilization, we random sample jobs to create simulation workloads. Also, the job processing time and node requirements are re-scaled to meet the size of the simulated data center.




\subsection{Methodology}

We evaluate the performance of the online algorithm under various types of workloads and various value of \emph{least service quality} $L$. We set the workload utilization range from $10\%$ to $150\%$. Note that as Random-Fit has its randomness factor internal to the algorithm, we do not need to tune its randomness. Each simulation is repeated for $30$ times and we compare the average values. To evaluate the algorithms, we conduct large scale simulations ($100$ machine nodes) to thoroughly compare the performance of the online algorithms. Due to the high running time demand of the offline algorithm, we simulate with relative smaller scale parameters ($16$ machine nodes) when compare the online algorithms with the optimal offline algorithm.


\subsection{Result and analysis}

We first present the evaluation of the online algorithms Random-Fit, First-Fit, Best-Fit and GreenSlot under various settings. Then we show the comparison of the online algorithms with the optimal offline algorithm to confirm with the theoretical competitive ratio analysis.

\subsubsection{Comparison of online algorithms}

We compare the online algorithm on the profits they achieve. In order to compare the competitive ratio of the online algorithms, we normalize the profits of each algorithm by the best-performed algorithm under each setting as detailed below.

First, we set the most profitable algorithm at each setting (under various workload utilizations and least service quality $L$ values) as an optimal performance $OPT'$. Then we compute the lower bound of competitive ratio using $OPT' / ALG$ where $ALG$ is the net profit gained by an online algorithm. As $OPT'$ is usually lower than the true optimum, therefore, the competitive ratio derived is only a lower bound of the real competitive ratios. It is fair enough to show that our designed  algorithm has better worst-case competitive ratios than First-Fit, Best-Fit, and GreenSlot.

\begin{figure}
  \centering
  \subfigure[Grid5k-1]{\includegraphics[width=.26\textwidth, clip = true, trim = {0mm 0mm 20mm 5mm}]{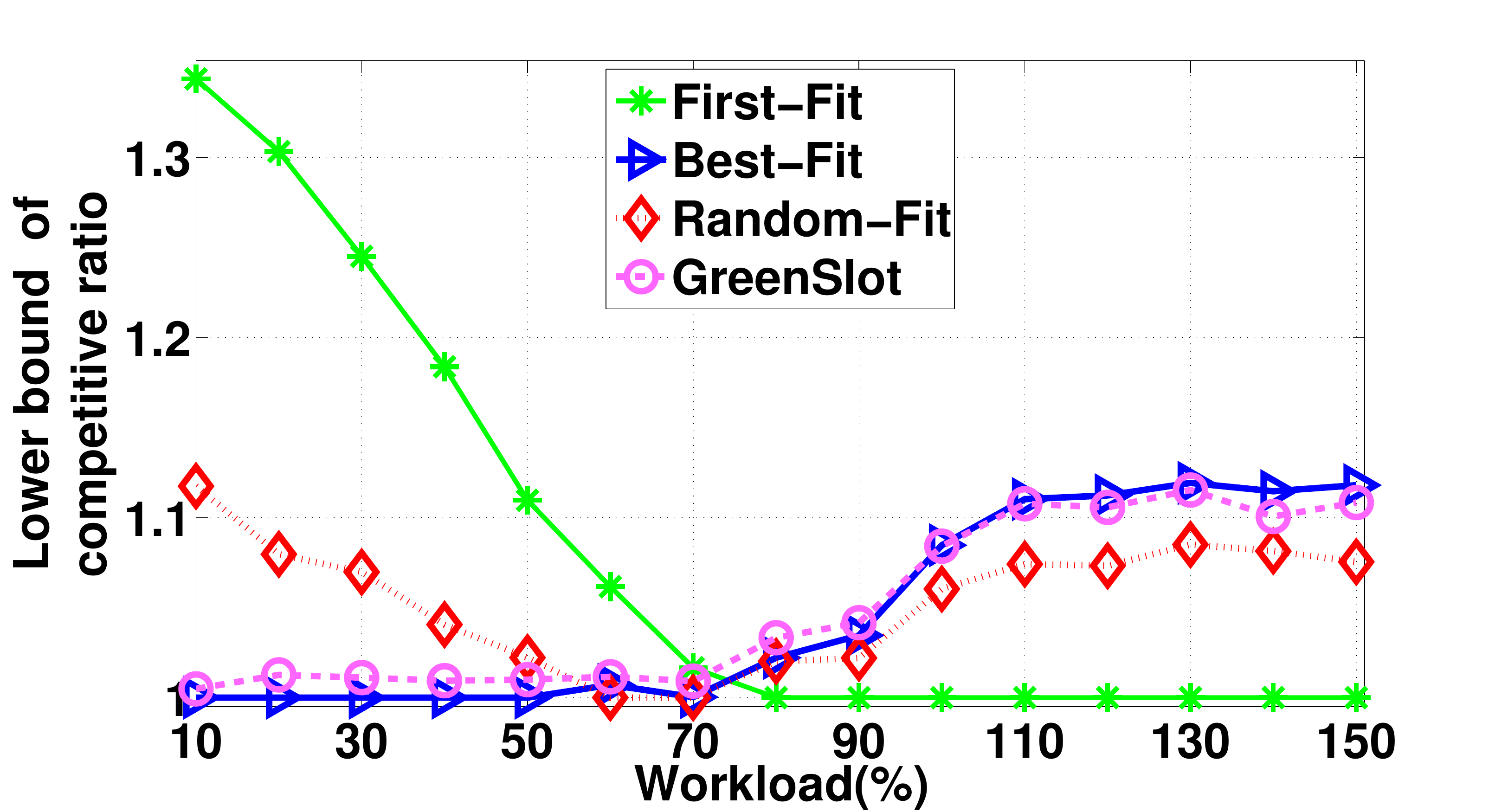}}
\subfigure[Grid5k-2]{\includegraphics[width=.26\textwidth, clip = true, trim = {0mm 0mm 20mm 5mm}]{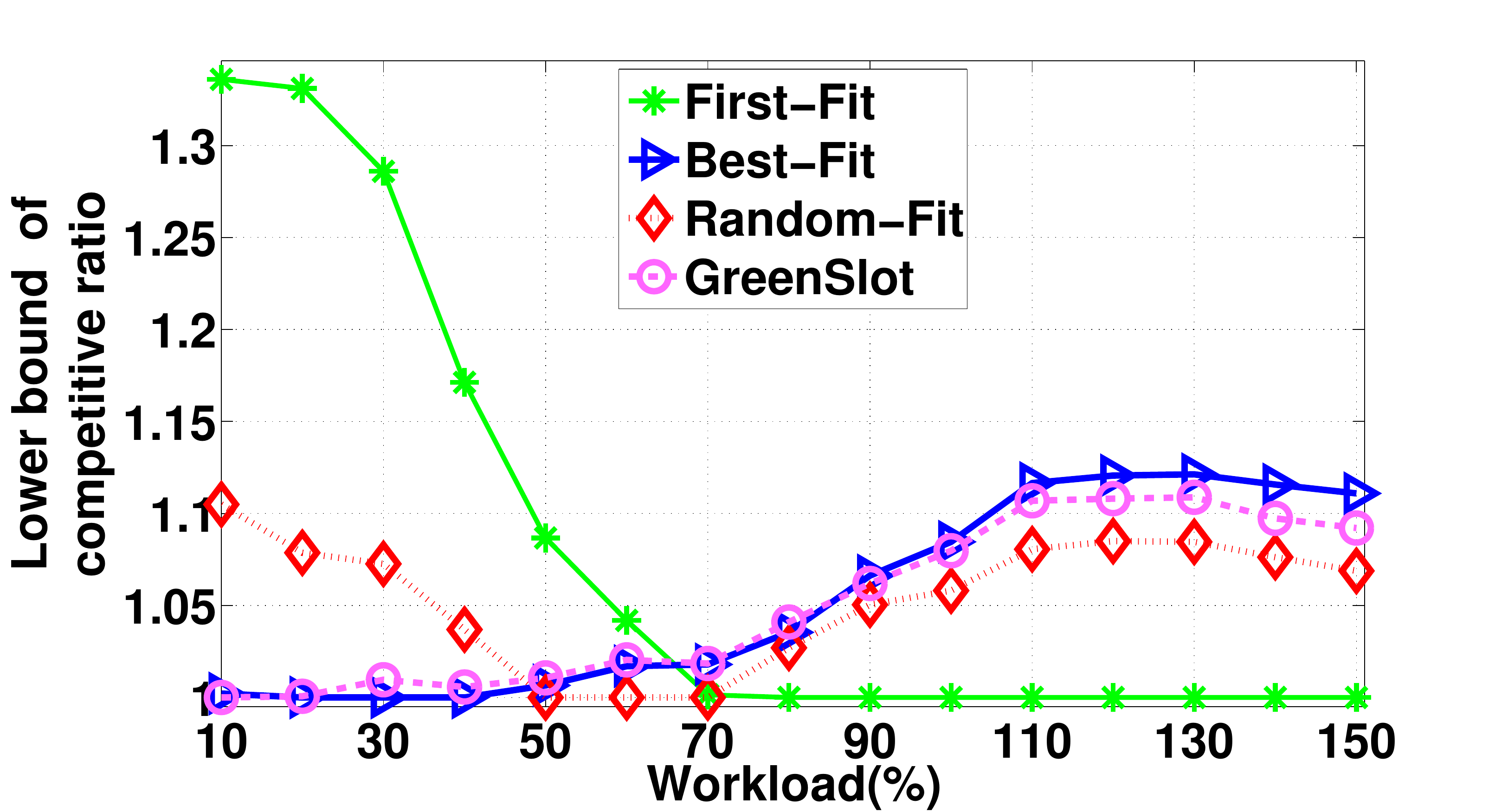}}
\caption{Lower bounds of competitive ratio  under different workloads \newline with $L = 0.2$}
\label{fig:prfits_0.2}
\end{figure}

\begin{figure}
  \centering
  \subfigure[Grid5k-1]{\includegraphics[width=.26\textwidth, clip = true, trim = {0mm 0mm 20mm 5mm}]{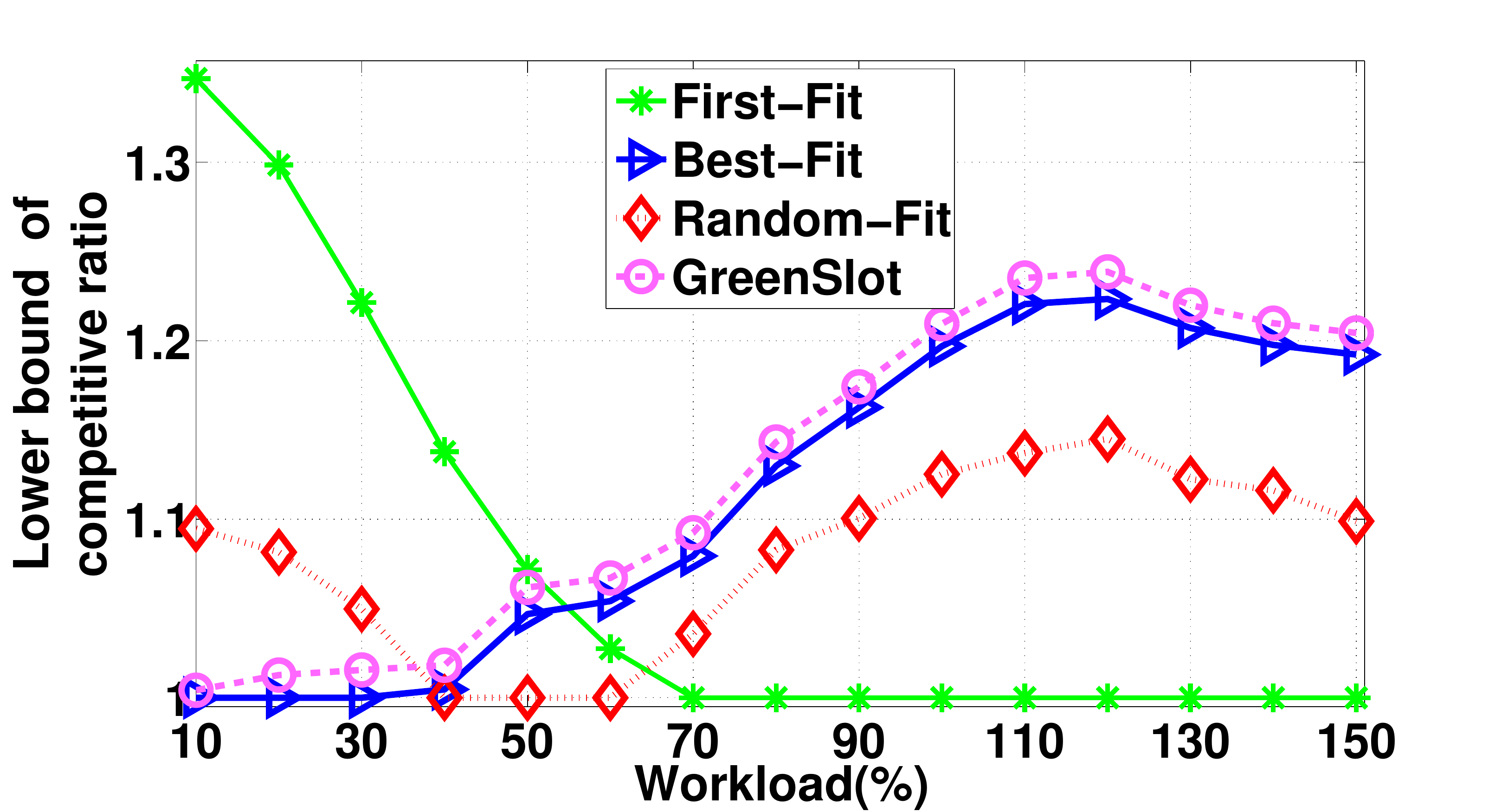}}
\subfigure[Grid5k-2]{\includegraphics[width=.26\textwidth, clip = true, trim = {0mm 0mm 20mm 5mm}]{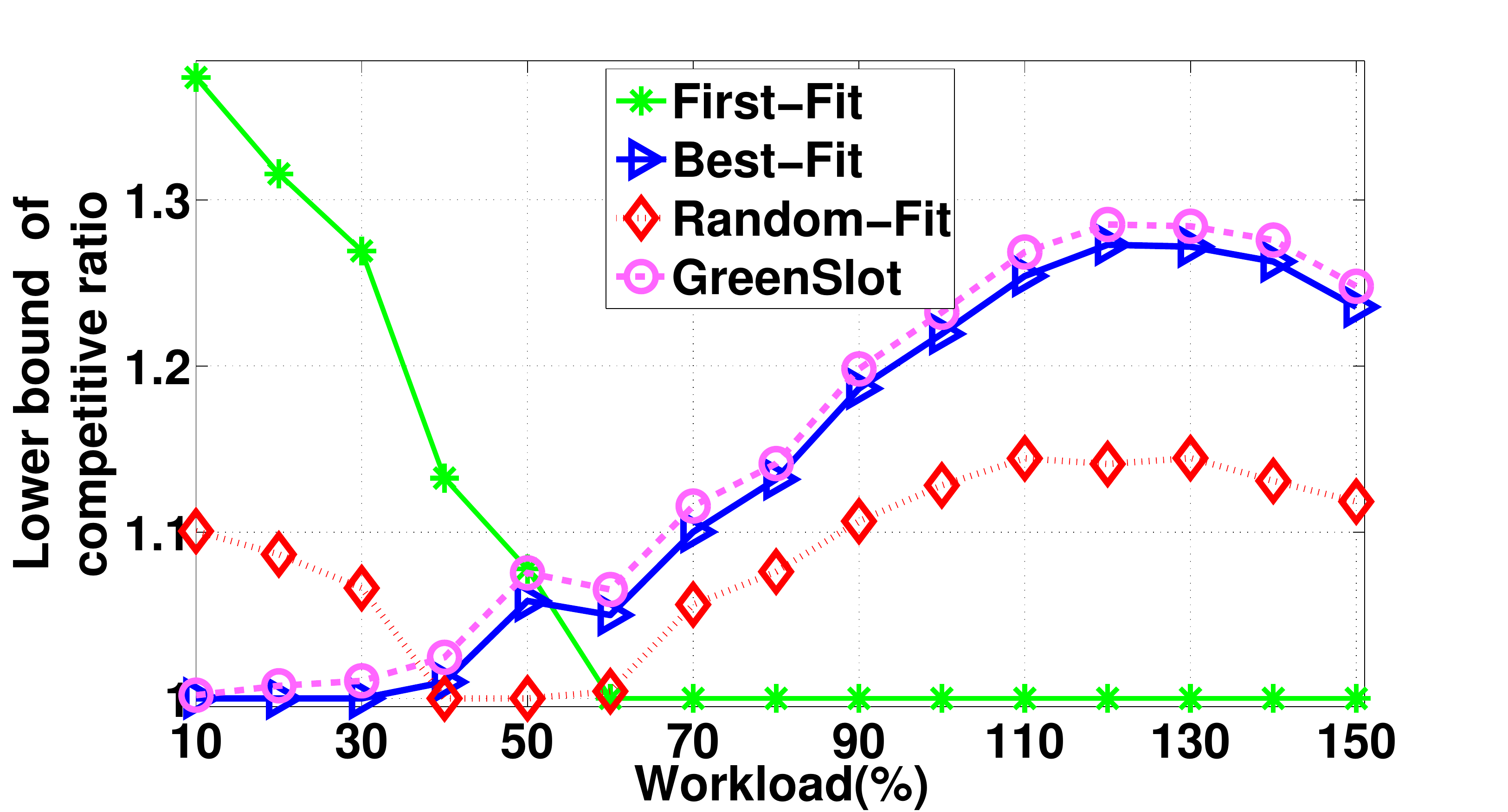}}
\caption{Lower bound of competitive ratio under different workloads \newline with $L = 0.05$}
\label{fig:prfits_0.05}
\end{figure}



Figure~\ref{fig:prfits_0.2} and Figure~\ref{fig:prfits_0.05} show the lowered bound of competitive ratios of the algorithms under various workloads settings and with least service quality $L$ as $0.2$ and $0.05$ respectively. From these figures, we observe that Best-Fit tends to gain a better profit when the data center utilization is lower than $60\%$, while First-Fit is better when the data center utilization is higher than about $80\%$. In whatever data center utilization, our proposed algorithms always guarantee a better worst-case performance. Note that an online algorithm cannot predict precisely a data center's long-time utilization at both fine-grained and coarse-grained levels. Therefore, alternatively employing the two algorithms First-Fit and Best-Fit cannot achieve a better worst-case performance than Random-Fit.

Best-Fit is less profitable when the data utilization is high because Best-Fit tends to delay scheduling jobs in order to consume less expensive energy. This delayed scheduling behavior results in many jobs missing their deadlines and thus achieving a lower profit. While First-Fit always schedules jobs to the first available time slots thus it could schedule more jobs than other algorithms. In the simulation, we observe First-Fit schedules $20\%$ more workloads then Best-Fit and GreenSlot, and around $10\%$ more workloads than Random-Fit with moderate workload (has utilization $70\%-100\%$). But it cannot make a good use of green energy when the data center is of low utilization. Its green energy utilization is less than $70\%$ of that of Best-Fit when workload utilization is around $10\%-60\%$. While Random-Fit can strike a balance between the amount of workload scheduled and the amount of green energy consumed, and thus tends to have better competitive ratio.

Taking the above analysis one step further, we conclude that if the data center utilization is predictable, then an adaptive scheduling algorithm which dynamically switches between Best-Fit and First-Fit according to the data center's utilization in a long-enough scheduling window would have a better performance. However, the data center utilization is usually hard to be predicted~\cite{MeisnerW10}.

In the simulation, we also find GreenSlot is sensitive to the value of the least service quality $L$. It has performance very close to Best-Fit when $L$ is relatively small, i.e., the job span is relatively large. It is because the penalty of delaying scheduling jobs will not be effective when the jobs have relatively small least service quality, as the penalty will be imposed only when a job is about to miss its deadline (for example, 20\% of its required processing time ahead of its deadline).

The running time of these algorithms in scheduling a job is in the order of several milliseconds which is negligible compare to the job's processing time, usually at several minutes or hours. In specific, First-Fit runs fastest, Random-Fit is the second, while GreenSlot and Best-Fit almost have the same running time.

Based on our simulation results, we remark that Random-Fit is the best algorithm (in terms of competitive ratio and profit maximization).

\subsubsection{Comparison with offline algorithm.}

We further conduct simulations to confirm with the theoretic result that Random-Fit has a better worst-case competitive ratio when jobs are of the same lengths and sizes. We implement an optimal offline algorithm to show the real experimental competitive ratios. The offline algorithm is formulated using a binary integer program and it is run by the LINDO solver. Note that the optimal algorithm is very time consuming, thus we shrink the nodes in the data center from $100$ to $16$ in order to get the optimal result within a reasonable time.

In the simulation, we simulate $2$ uniform workloads with utilization $10\%$ and $100\%$ respectively. We compare the online algorithms against the optimal offline algorithm using competitive ratio. For ease of presentation, we abbreviate the algorithms First-Fit, Best-Fit, Random-Fit, GreenSlot and offline optimal as: FF, BF, RF, GS and OPT respectively.


\begin{table}[!ht]
\centering
\caption{competitive ratio of online algorithms}
\begin{tabular}{|l|l|l|l|l|l|l|}  \hline
matrix & FF & BF & RF & GS \\ \hline \hline
competitive ratio (workload = 10\%) &1.56&	1.03&	1.16&	1.03\\ \hline
competitive ratio (workload = 100\%)      &   1.05 &   1.29   & 1.24 &   1.27   \\ \hline
\end{tabular}
\label{tb_com_opt_20}
\end{table}


Table~\ref{tb_com_opt_20} shows the competitive ratio of various online algorithm under different workload utilization. We conclude that First-Fit, Best-Fit and GreenSlot have competitive ratios worse than the theoretical upper bound ($1.25$) of Random-Fit. This conclusion confirms our theoretical results.


\section{Conclusions}

In this work, we study online scheduling of energy and jobs in green data centers with the objective of maximizing net profit. In our problem setting, energy costs are time-sensitive and so is the net profit. Prior work employs deterministic approaches only and the underlying algorithmic ideas are either First-Fit or Best-Fit; furthermore no theoretical analysis has been given. In this paper, competitive analysis is used to measure an online algorithm's theoretical performance. We conclude that randomness plays an important role in maximizing net profit. Experiments on real workload traces have shown that our algorithm indeed outperforms the previous ones, as what the theory indicates.


\bibliographystyle{IEEEtran}
\bibliography{greenSlot-short}


\appendix
\renewcommand{\thesubsection}{\Alph{subsection}}


\subsection{Offline Algorithm}
\label{Appendix_offline}
We formulate a linear program for the special cases when jobs have same processing times and node requirements. Let $g(t)$ denotes the amount of green energy arrive at time $t$ and let $b(t)$ denotes the unit brown energy price at time $t$. Assume all jobs have the same processing time slots $p$ and node requirement $q$. Let $v_j$ denote the revenue earned by scheduling job $j$. Let $y_j$ be an indicator variable indicates whether a job is scheduled ($y_j = 1$) or not ($y_j = 0$). Let $s[j, t]$ be an indicator variable denotes whether job $j$ is started at time $t$ ($s[j, t]=1$) or not ($s[j, t] = 0$). Let $n(t)$ denotes the number of jobs started at time $t$. Let $e(t)$ denotes the energy demand at time $t$.

We have the following formulation.

\begin{align*}
\max & R - E &\\
\mbox{subject to } & R = \sum_{j \in J} v_j \cdot y_j &\\
& E = \sum^T_{t = 1} \max\{0, e(t) - g(t)\} \cdot b(t) &\\
& n(t) = \sum^J_{j = 1} s[j, t] & \forall j\\
& e(t) = \sum_{\max\{0, t - p + 1\} \le k \le t} n(k) \cdot q  & \forall t\\
& e(t) \le M & \forall t\\
& \sum^T_{t = 1} s[j, t] \ge y_j & \forall j\\
& \sum^T_{t > d_j} s[j, t] = 0 & \forall j\\
& \sum^T_{t < r_j} s[j, t] = 0 & \forall j\\
& s[j, t] = \{0, 1\} & \forall j, t\\
& y_j = \{0, 1\} & \forall j
\end{align*}

\subsection{Hardness of the Problem GDC-RM}
\label{appendix_GDC-RM_hardness}
Note that GDC-RM essentially is not an offline problem since the jobs and the green energy cannot be modeled and predicted precisely at all the time. However, understanding the hardness of the offline version may be useful to us in evaluating an online algorithm's theoretical and empirical performance. We prove that the offline version of GDC-RM is NP-hard, using a reduction from the well-known NP-hard problem `Knapsack'~\cite{GareyJ79}.

\begin{theorem}
The offline version of the problem GDC-RM is NP-hard.
\label{thm:nphard}
\end{theorem}

\begin{IEEEproof}
Given a candidate solution, it takes polynomial-time for us to verify whether this solution is feasibly scheduled or not. Thus, the problem GDC-RM belongs to NP. In the following, we prove that GDC-RM is NP-hard by showing a polynomial-time reduction from the Knapsack problem to it. In the Knapsack problem, there are a knapsack of capacity $W$ and $n$ items with each one has size $s_i$. The goal is to make the knapsack as full as possible. The Knapsack problem is known NP-hard~\cite{GareyJ79}.

Consider the problem GDC-RM. Assume the produced green energy has a budget of $B$ in a scheduling window and the brown energy's costs ($B^d$ and $B^n$) are high enough such that any use of brown energy makes no positive net profit at all. Therefore, to maximize the net profit, we would like to find a set of jobs such that these jobs consume as much as close to but no more than the green energy budget $B$ without using any amount of the brown energy. Particularly, we restrict that the green energy is available within a scheduling window $[t, t']$ and all jobs $j$ have the same release time $t$ and (maybe different) deadlines $d_j$ $\left(:= t + \frac{p_j}{L} \le t'\right)$ to ensure the same service qualities $L$ --- $t$ and $t'$ are the boundaries of this scheduling window and $t'$ is the latest time where green energy is still available. Let $t' - t = W$. Also, we restrict that each job $j$ has $q_j = 1$. This conversion takes linear time of the number of jobs.

If we have a polynomial-time optimal solution to the problem GDC-RM with the special input instance as created as in the above, then we have an optimal solution to the following Knapsack problem: The knapsack has its capacity of $W = t' - t$ and each item $j$ has its size of $p_j$. As the Knapsack problem is NP-hard, then the problem GDC-RM is NP-hard.
\end{IEEEproof}




\end{document}